\documentstyle[preprint,aps]{revtex}
\begin{document}
\draft
\title{
Optical Conductivity in Mott-Hubbard Systems
}

\author{M. J. Rozenberg\cite{new}, G. Kotliar, and H. Kajueter}
\address{
Serin Physics Laboratory, Rutgers University,
Piscataway, NJ 08855-0849
}
\author{G. A. Thomas and D. H. Rapkine}
\address{
AT\&T Bell Laboratories, Murray Hill, NJ 07974-0636
}
\author{J. M. Honig and P. Metcalf}
\address{
Department of Chemistry, Purdue University, West Lafayette, IN 47907
}
\maketitle

\begin{abstract}

We study the  transfer of spectral weight in the
optical spectra of a strongly  correlated
electron system as a function of temperature and interaction strength.
Within a dynamical mean field theory
of the Hubbard model
that becomes exact in the limit of large
lattice coordination,
 we predict an anomalous enhancement of spectral weight
as a function of temperature in the correlated metallic state and report on
experimental measurements which agree with this prediction
in $V_2O_3$. We argue that the
 optical conductivity anomalies in the metal are connected to
the proximity to a crossover region in the phase diagram
of the
model.
\end{abstract}

The interest in the distribution of spectral weight in the
optical conductivity of correlated electron systems has been revived by
the improvement in the quality of the experimental data
\cite{thomas,bucher}.
The traditional theoretical
methods used in the strong correlation problem,
have only been partially successful in describing the interesting
behavior of the optical response which takes place
in the strong correlation regime.
This is most notable when the temperature dependence of the spectral
weight is considered.
We present here new optical conductivity data on metallic
$V_2 O_3$ and analyze  the problem theoretically
it using  a mean field theory which is
exact in the
limit of large number of spatial dimensions \cite{vollmetz}.
In this limit, lattice models
can be mapped
onto an equivalent impurity model subject to a selfconsistency
condition for the conduction electron bath \cite{gk,gkq}.
This technique has already  given  some new insights into the Mott transition,
which is realized in the solution of the Hubbard model
\cite{motthubb,pcj,gekr,rkz}.
The new  experimental measurements on $ V_2O_3 $
test a critical prediction of the theory.

The goal of this  paper is  to demonstrate
two  points: {\it i)} that
the present dynamical mean field approach is {\it a useful tool}
which allows
us to make qualitative predictions such as an anomalous
enhancement of spectral weight  as a function of  temperature.
This behavior of the spectral weight is confirmed by  experiments in
$V_2 O_3$ reported here;
and {\it ii)} that  the Hubbard model on a {\it frustrated } lattice,
in the limit of large lattice coordination, with parameters
extracted from the optical data, accounts  for many other
experimental features of  the  $ V_2 O_3$ system,
such as its phase diagram and large specific heat capacity.
We also present results for the various quantities which are useful
in the analysis of the optical conductivity of the Hubbard model
, treated in  an   approximation which is exact
in a well defined limit.

The experiments were carried out on single crystals of
$V_2 O_3$ which were grown by the skull-melting process,
carefully annealed, polished and mounted on a temperature-controlled
stage in an optical cryostat \cite{thomas,carter}.
The specular reflectivity was measured using a Fourier transform
spectrometer over a frequency range from $6 meV$ to $3.7 eV$.
The data were combined with dc conductivity and higher energy
reflectivity measurements, and a Kramers-Kronig
transformation was used to determine the conductivity.

The theoretical model is the Hubbard hamiltonian,
\begin{equation}
H = -\sum_{<i,j>} (t_{ij}+\mu) c_{i\sigma}^{\dagger} c_{j\sigma}
 +  \sum_i U (n_{i \uparrow}-\frac{1}{2}) (n_{i\downarrow}-\frac{1}{2})
\label{HubHam}
\end{equation}
where summation over repeated spin indices is assumed.

In the  limit of infinite dimensions the coordination number
of the lattice  (i.e. the number of neighbors $q$) gets large and
the hopping is scaled as
$t_{ij} \rightarrow \frac{t}{\sqrt{q}}$ \cite{vollmetz}.
The lattice model is mapped onto an
equivalent impurity problem supplemented by a selfconsistency condition.
The derivation of this result can be found  elsewhere \cite{gk,gkq},
so we only present
the final expressions.
The resulting local effective action reads,
\begin{equation}
{S}_{local}=
- \int^{\beta}_{0} d\tau \int^{\beta}_{0} d\tau'
c^{\dagger}_{\sigma}(\tau)  G_{0}^{-1}(\tau-\tau') c_{\sigma}(\tau')
\ + \  U \ \int^{\beta}_{0}d\tau (n_{\uparrow}(\tau)-{1\over2})
(n_{\downarrow}(\tau)-{1\over2})
\label{localaction}
\end{equation}
where $c^\dagger_\sigma,\ c_\sigma$ correspond to a particular site.
Requiring that $G_{local}(\omega)=\Sigma_k G(k,\omega)$
the selfconsistency condition becomes
$
 G_{0}^{-1}(\omega)= \omega + \mu - t^2 G_{local}(\omega)
$,
where we have assumed a semi-circular bare
density of states
$\rho^o (\epsilon) =
(2 / {\pi D}) \sqrt{1 - ( \epsilon /D)^2}$, with $t=\frac{D}{2}$,
which can be realized in a Bethe lattice
and also on a fully connected fully frustrated version of
the model \cite{zrk}.
We consider the symmetric case $\mu = 0$.
We use an exact diagonalization algorithm (ED) \cite{ed}
and an extension of the  second order perturbative (2OPT) calculation
to solve the associated
impurity problem \cite{zrk}.

Early work on the optical conductivity in the Hubbard model
was carried out by Pruschke {\sl et al.} using
quantum Monte Carlo \cite{pcj}, a technique that is restricted to
high temperatures.
The 2OPT and ED techniques allow us to study the low temperature regime
near the Mott transition relevant to the experiments addressed here.

The optical conductivity is defined as
$\sigma(\omega) = -\frac{1}{\omega} {\rm Im} \langle [j,j] \rangle$.
In the limit of $d=\infty$ it can be expressed in terms of the
one particle spectrum \cite{pcj,khurana},
\begin{equation}
\sigma(\omega)= \\
\frac{1}{\omega} \frac{2 e^2 t^2 a^2}{\nu \hbar^2}
\int_{-\infty}^{\infty}{d\epsilon}\ {\rho^o(\epsilon)}
\int_{-\infty}^{\infty}\frac{d\omega'}{2\pi}\
A_\epsilon(\omega') A_\epsilon(\omega'+\omega)
(n_f(\omega')-n_f(\omega'+\omega))
\label{conv}
\end{equation}
with $A_\epsilon(\omega) = -2 {\rm Im}[G_k(\omega)]$
, $e$ the electron charge, $\nu$ the volume of
the unit cell, and $a$ the lattice constant.

At $T=0$,  $\sigma(\omega)$
can be parametrized by
$\sigma(\omega)=
\frac{{\omega^*_P}^2}{4\pi}\delta(\omega) + \sigma_{reg}(\omega) $
where the coefficient in front of
the $\delta$-function is the Drude weight and
$\omega^*_P$ is the renormalized plasma frequency \cite{kohn}.
In the presence of disorder
$\delta(\omega)$ is replaced by a lorentzian of width $\Gamma$.
The expectation value of the
kinetic energy $\langle K \rangle$ is related to the
conductivity by the sum rule
\begin{equation}
\int_0^\infty\sigma(\omega) d\omega = -\frac{\pi e^2 a^2}{2d\hbar^2 \nu}
\langle K  \rangle =
\frac{\omega_P^2}{4\pi}
\label{sum}
\end{equation}
In the  limit $d \rightarrow \infty$ the Drude weight
is obtained in terms of the quasiparticle weight $Z$,
$\frac{{\omega^*_P}^2}{4\pi} = \frac{4\pi t^2 e^2 a^2}{\hbar^2\nu}Z\rho^o(0)$.

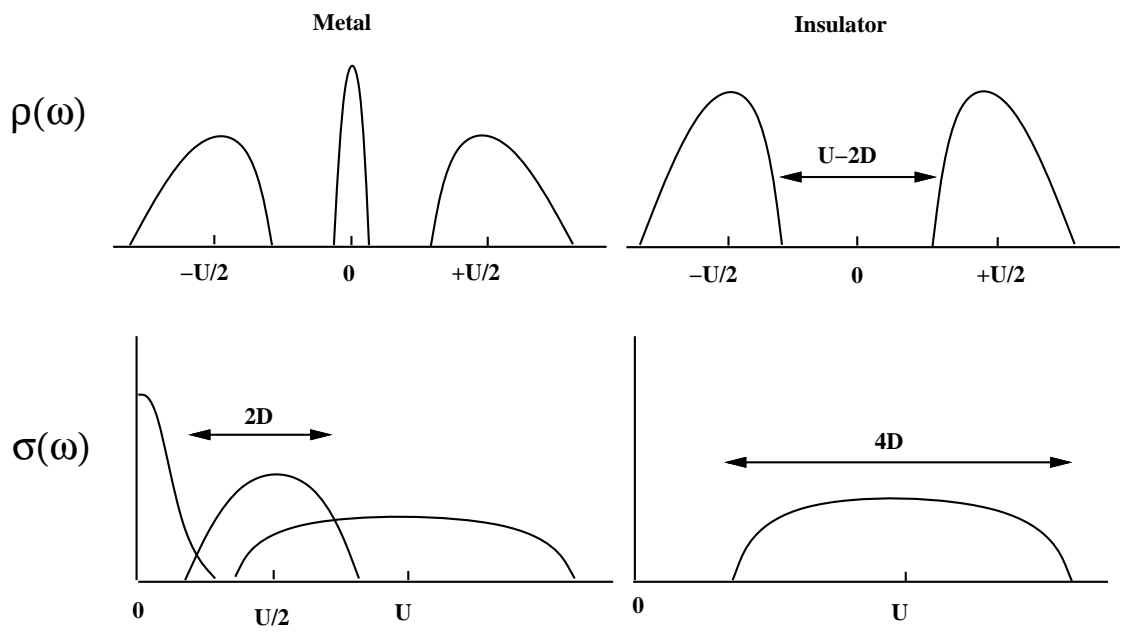
\begin{figure}
\caption{Schematic DOS (1/2 filling) and their
corresponding optical spectra for the metallic and insulator solutions.
The approximate width of the incoherent peaks in the $DOS$ is $2D$ and that
of the central peak in the metal is $ZD \equiv \epsilon_F^*$.
}
\label{figa}
\end{figure}

The solution of the mean field equations shows that at low temperatures the
model has a metal insulator transition (Mott-Hubbard transition) at an
intermediate value of the interaction $U_c \approx 3D$
\cite{motthubb,pcj,gekr,rkz}.
The metallic and insulator solutions at low $T$ are very different and we
schematically represent them in Fig. \ref{figa}.
The optical conductivity response, in a first approximation,
can be understood from transitions between the states which
appear in the
the $DOS$.
We schematically sketch the optical response  in the lower part of
Fig. \ref{figa}.

We will discuss the results for the model in regard of
different experimental data on the $V_2 O_3$ system.
Vanadium oxide has three
$t_{2g}$ orbitals per $V$ atom which are filled with two electrons.
Two electrons (one per $V$) are engaged in a strong cation-cation bond,
leaving the remaining two in a twofold degenerate
$e_g$ band \cite{castellani}.
LDA calculations give a bandwidth of $\sim 0.5eV$  \cite{lda}.
The Hubbard model ignores the degeneracy of the band which is
crucial in understanding the magnetic structure \cite{castellani}, but
captures the interplay of the electron-electron
interactions and the kinetic energy.
This delicate interplay of itinerancy and localization is responsible for
many of the anomalous properties of this compound, and are correctly predicted
by this simplified model.

Experimentally one can vary the ratio $U \over D$ by introducing $V$
vacancies.
The parameters are extracted from the experimental data
on $V_{2-y} O_3$
by comparing the measurments with the schematic spectra
displayed in Fig. \ref{figa}, and
are summarized in table \ref{table1}.
It is not surprising that $U$ and $D$ are different in the metal and the
insulator, since the lattice parameter and the screening length change rapidly
across the phase boundary.

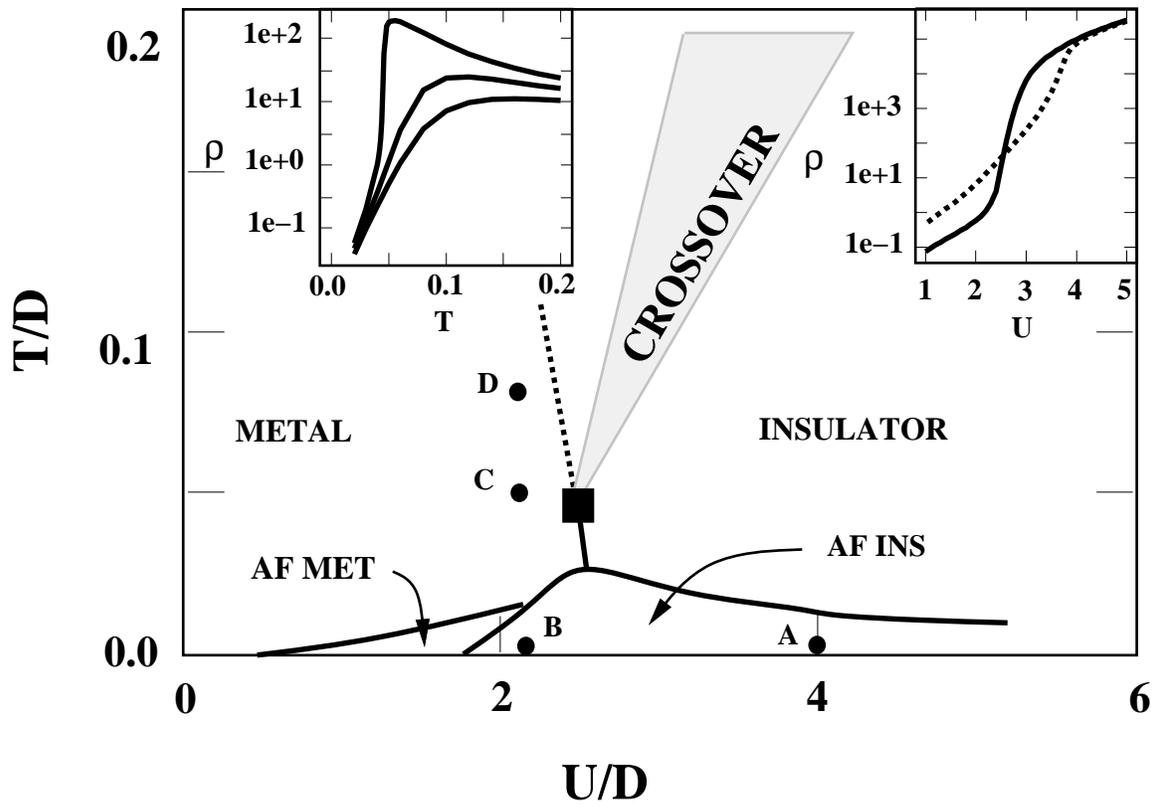
\begin{figure}
\caption{Approximate phase diagram for the model with $n.n.$ and
$n.n.n.$ hopping
$(t_2/t_1)=\protect\sqrt{1/3}$.
The $1st$ order PM metal-insulator
transition ends at the critical point $T_{MIT}$ (square).
The dotted line and the shaded region describe two
crossovers as  discussed
in the text.
The full circles indicate the position
of the optical spectra. $A$: insulator
($y=0$), $B$: insulator ($y=.013$), $C$: metal ($y=0,\ 170K$),
$D$: metal ($y=0,\ 300K$).
Note that for comparison with experimental results increasing $U/D$
is associated with decreasing pressure \protect\cite{kuwamoto,mcwhan}.
Left inset:
$\rho_{dc}(T)$ for $U/D=2.1,2.3,2.5$ (bottom to top).
The maxima of $\rho_{dc}(T)$ defines the dotted line.
Right inset: $\rho_{dc}(U)$ for $T=0.06D$ (full) and $T=0.15D$ (dotted).
}
\label{figb}
\end{figure}

Our calculations have been performed on
the model
with nearest neighbor hopping
$t_1 \over {\sqrt{q}}$ and
next to nearest neighbor hopping $t_2 \over q$
on a Behte lattice.
The ${n.n.n.}$ hopping introduces  magnetic frustration which
is essential if we want to describe $V_2 O_3$ with a one band model.
The condition $t_1^2 + t_2^2 = t^2$ keeps the
bare density of states $\rho^0$ invariant. For $t_2/t_1 =0$ we recover the
original hamiltonian, and $t_2/t_1 =1$ gives the PM solution.
The same model on the hypercubic lattice
would give very similar results.
In Fig. \ref{figb} we display the phase diagram
for $t_2/t_1 =\sqrt{1/3}$. It  has the same topology
as the experimental one \cite{carter,kuwamoto,mcwhan}.
Frustration lowers the $T_{Neel}$ well below  the second order
 $T_{MIT}$ point \cite{rkz}.
Using the parameters of table \ref{table1},
$T_{MIT}\approx 240K$,
which is only within less than a factor of 2 from the experimental result.
The dotted line  indicates a crossover separating
a good metal at low $T$ and a semiconductor  at higher $T$.
Between these states
$\rho_{dc}$ has an anomalous rapid increase with $T$.
The reason for this feature
can be traced to the disappearance of the coherent central quasiparticle
peak in the $DOS$.  We find the behavior of $\rho_{dc}$ to be in good
agreement with the
experimental results of Mc Whan {\sl et al.} (inset Fig. \ref{figb})
\cite{mcwhan}.
Another crossover is indicated by a shaded area, it
separates a semiconducting region with a gap
$\Delta$ comparable with $T$ from a good insulator where $T << \Delta$,
consequently the crossover temperature  increases  linearly
with $U$ and the crossover width  becomes broader with increasing $T$.
 This crossover is characterized
by a sudden increase in $\rho_{dc}$ as function of
$U$ at a fix $T$ (inset Fig. \ref{figb}).
This crossover was experimentally
observed in $V_2 O_3$ by Kuwamoto {\sl et al.}
\cite{kuwamoto}.

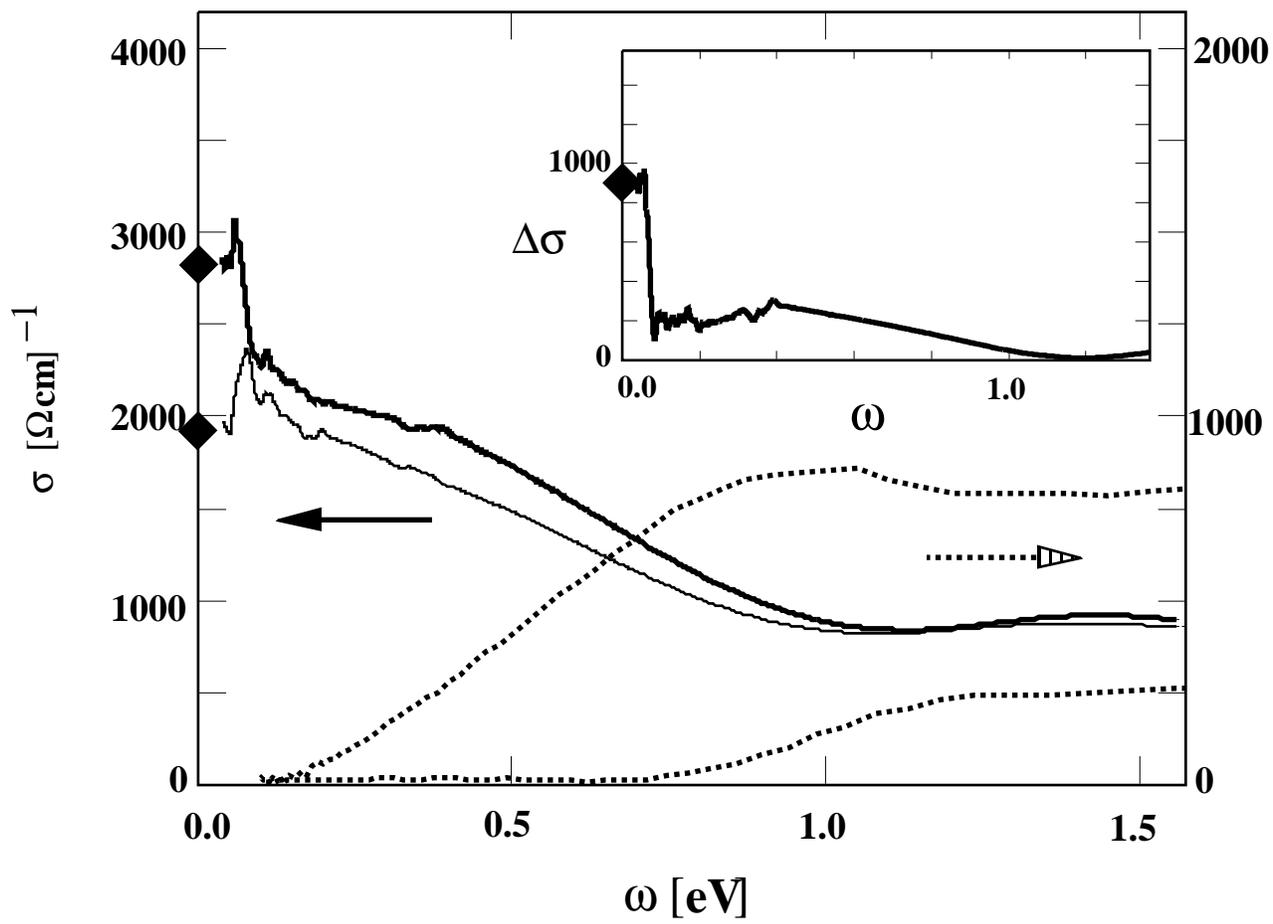
\begin{figure}
\caption{The experimental $\sigma(\omega)$ of metallic $V_2O_3$ (full lines)
at $T=170K$ (upper) and $T=300K$ (lower). The inset contains the difference
of the two spectra
$\Delta\sigma(\omega)=\sigma_{170K}(\omega)-\sigma_{300K}(\omega)$.
Diamonds indicate the measured dc conductivity $\sigma_{dc}$.
Dotted lines indicate $\sigma(\omega)$ of insulating
$V_{2-y}O_3$ with $y=.013$ at $10K$ (upper) and $y=0$ at $70K$ (lower).
}
\label{figc}
\end{figure}

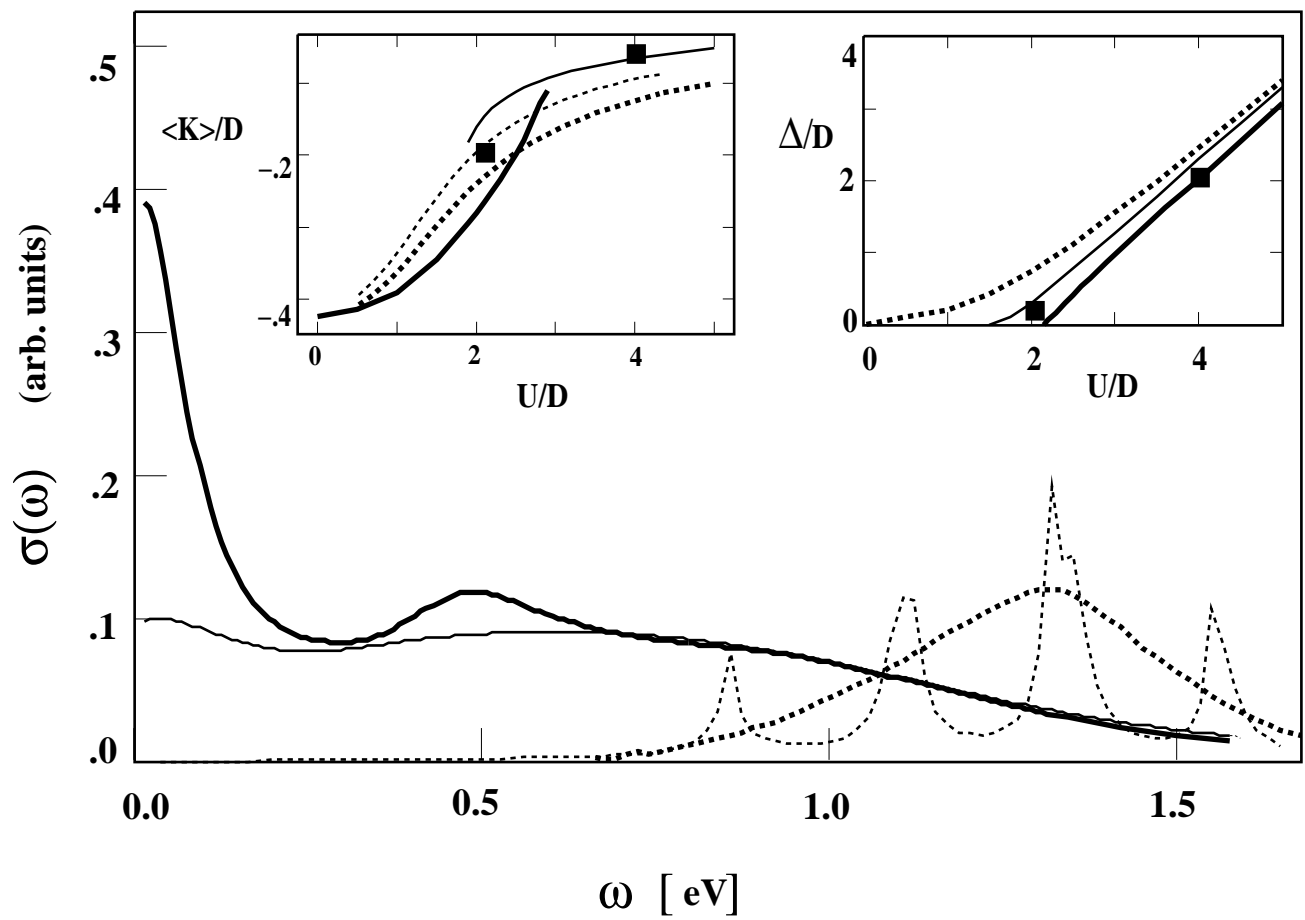
\begin{figure}
\caption{$\sigma(\omega)$ for the metallic solution (full lines)
at $U=2.1D$ and $T=0.05D$ (upper) and $0.083D$ (lower).
A small $\Gamma = 0.3$ and $0.5D$ was included to
mimic a finite amount of
disorder. Dotted lines indicate the insulating
solution results at $U=4D$ and $T=0$ from ED (thin-dotted) and 2OPT
(bold-dotted). The peaks in ED result
from the finite size of the clusters that can be considered.
Left inset: $\langle K \rangle$ versus
$U$ for the AFM (bold-dotted) and PM insulators (thin), PM metal (bold)
and partially frustrated model (thin-dotted).
Right inset: $\Delta$ versus $U$ for
the AFM (dotted), partially frustrated (thin) and PM
(bold) insulators. $\Delta$ is twice the energy of the
lowest pole from the ED Green function.
The data are for $N_{sites} \rightarrow \infty$
from clusters of 3, 5 and 7 sites assuming ${1}/{N_{sites}}$
scaling behavior.
Black squares show the insulator experimental results.
}
\label{figd}
\end{figure}

The experimental optical
spectrum of the  insulator is
reproduced in Fig. \ref{figc} (dotted lines).
It is characterized by an excitation gap at
low energies, followed by an incoherent feature that corresponds to
charge excitations of mainly Vanadium character \cite{thomas}.
These data are to be compared with the model results of Fig. \ref{figd}.
The size of the gaps $\Delta$
(shown for various degrees of
magnetic frustration in the inset of Fig. \ref{figd}),
and the overall shape of the
spectrum are found to be in good agreement with the
experimental results \cite{thomas}.
Another quantity that can be compared to the experiment
is the integrated spectral
weight $\frac{\omega_P^2}{4\pi}$ which
is related to $\langle K  \rangle$ by the sum rule (\ref{sum}).
Setting the lattice constant $a\approx 3 \AA$ the average $V-V$ distance,
we find our results also in good agreement with the experiment
(left inset of Fig. \ref{figd}). Notice that the dependence
of $\langle K  \rangle$
and $\Delta$ on the degree of magnetic frustration is
not very strong, which is consistent with the view that magnetism
in this system is not the driving mechanism of the metal insulator transition.
Notice also that the experimental points are closer to the
intermediate frustration
curves than to the unfrustrated ones.

We now  discuss the new data in  the  metallic phase.
The experimental data
for pure samples that become insulating at $T_c \approx 150K$
were obtained for $T =170K$ and $T=300K$, and are presented in
Fig. \ref{figc} (full lines).
Both spectra  are made up of broad absorption at higher frequencies
 and some phonon lines in the far infrared.
They appear to be rather
featureless, however, upon considering their difference
(in which the phonons are approximately eliminated)
distinct features are observed.
As $T$ is lowered, there is
an enhancement of the spectrum at intermediate frequencies of
order $0.5eV$; and more notably, a sharp low
frequency feature emerges that extends from $0$ to $0.15eV$.
Moreover, these enhancements result in an anomalous {\it change} of
the total spectral weight $\frac{\omega_P^2}{4\pi}$ with $T$.
We argue below, that these observations can be accounted by
the Hubbard model treated in mean field theory.

In Fig. \ref{figd} we show the calculated  optical spectra
obtained from 2OPT for two different values of $T$.
The interaction is set to $U=2.1D$ that places the system
in the correlated metallic state.
It is clear that,
at least, the qualitative aspect of
the physics is already captured.
As the temperature is lowered, we
observe the enhancement of the incoherent structures at
intermediate frequencies of the order $\frac{U}{2}$ to $U$
and the rapid emergence
of a feature at the lower end of the spectrum.
Setting $D \approx 0.4eV$ we find these results
consistent with the experimental data on $V_2 O_3$.
The two emerging features can be interpreted from the qualitative
picture that was discussed before
which is relevant for low $T$.

An interesting prediction of the model is the anomalous increase
of the integrated spectral
weight $\frac{\omega_P^2}{4\pi}$ as $T$ is decreased.
This is due to the rather strong $T$ dependence of the kinetic
energy $\langle K \rangle \propto \frac{\omega_P^2}{4\pi}$
in the region near the crossover indicated by a dotted
line in the phase diagram.
This also corresponds with the behavior of the low frequency features in the
spectra.
{}From the model calculations  we expect an
enhancement of the spectral weight
as $T$ is decreased.
With the chosen parameters
this should occur at a scale
$T_{coh} \approx 0.05D \approx 240K$
which correlates well with the experimental data.
$T_{coh}$ has the physical meaning of the
temperature below which the Fermi
liquid description applies \cite{rkz},
as the
quasiparticle resonance emerged in the density of states.
We find $\frac{\omega_P^2}{4\pi} \approx
1000 \frac{ev}{\Omega cm}$ which is somewhat lower than the experimental
result.
This could be due to the contribution
from tails of bands at higher energies that are not
included in our model, or it may indicate that the
bands near the Fermi level are degenerate.

Experiment show that  the slope of the linear term in the
specific heat $\gamma$  in the metallic phase
is  unusually  large.
For $0.08\  Ti$ substitution $\gamma \approx 40 \frac{mJ}{mol K^2}$, while
for a pressure of $25 Kbar$ in the pure compound
$\gamma \approx 30 \frac{mJ}{mol K^2}$ and  with $V$ deficiency in
a range of $y=0.013$ to $0.033$ the value is
$\gamma \approx 47 \frac{mJ}{mol K^2}$  \cite{gamma}.
In our model $\gamma$ is simply related to the weight in
the Drude peak in the optical conductivity and to the quasiparticle
residue $Z$ ,
$\gamma = \frac{1}{ZD}3 \frac{mJ eV}{mol K^2} $.
The values of  $U=2.1D$ and
$D \approx 0.4eV$
 extracted from the optical data
correspond to a quasiparticle residue $Z \approx 0.3$,
and results in $\gamma \approx  25  \frac{mJ}{mol K^2}$ which
is close to
the experimental findings.
Thus, it turns out that the mean field
theory of the Mott transition
explains in a natural and qualitative manner, the experimentally observed
 optical conductivity spectrum,  the
anomalously large values of the slope of the specific heat $\gamma$,
and the dc conductivity in the metallic phase
as consequence of the appearance of a single small energy scale,
the renormalized Fermi energy $\epsilon^*_F$.

To conclude,
we have illustrated how the mean field
theory, that becomes exact in the limit of large dimensions, can be
used to  study the physics of systems where the
interactions  play a major role.
We presented new data on the metallic phase of  $V_2 O_3$ revealing
an anomalous enhancement of  the optical spectral weight
as the temperature is lowered in agreement with the surprising
prediction of the mean field theory.
We also  contrasted several experimental results to the solution of the
model Hamiltonian to argue that, in the mean field approximation
the Hubbard model gives a consistent semiquantitative picture
of this strongly correlated electron system.
A great challenge is to extend the mean field approach
to incorporate more realistic band
structure density of states with orbital degeneracies
and more complicated unit cells.
This extensions will require more efficient tools for
solving larger impurity models, and   would allow for a more
quantitative description of the  physical properties of
transition metal oxide  systems.

We acknowledge valuable discussions with V. Dobrosavljevic,
G. Moeller A. Ruckenstein  Q. Si and C.M. Varma.
This work was supported by the NSF
grant DMR 92-24000

\begin{table}
\begin{center}
\begin{tabular}{lcccc}
Phase & \multicolumn{4}{c}{Parameter }  \\
  &D [eV] &U [eV] &$\Delta$ [eV] & $\protect \omega_P^2/4\pi$ [eV/$\Omega$cm]
 \\
 \tableline
Insulator (y=0)  & $0.33 \pm 0.05$ & $1.3 \pm 0.05$ & $0.64 \pm 0.05$ &
$170 \pm 20$  \\
Insulator (y=.013) & $0.46 \pm 0.05$ & $0.98 \pm 0.05$ & $0.08 \pm 0.05$ &
$800 \pm 50$ \\
Metal (170K)  &$ 0.4  \pm 0.1$ &$ 0.8 \pm 0.1$ & -- &
$ 1700 \pm 300$ \\
\end{tabular}
\caption{Experimental parameters for the model.}
\label{table1}
\end{center}
\end{table}

\end{document}